# Giant coercivity induced by perpendicular anisotropy in $Mn_{2.42}Fe_{0.58}Sn$ single crystals

Weihao Shen[1], Yalei Huang[1], Xinyu Yao[1], Fangyi Qi[1], Guixin Cao[1,2,*]

[1]Materials Genome Institute, Shanghai University, 200444 Shanghai, China

[2]Zhejiang Laboratory, Hangzhou 311100, China

## Abstract

We report the discovery of a giant out-of-plane coercivity in the Fe-doped $Mn_3Sn$ single crystals. The compound of $Mn_{2.42}Fe_{0.58}Sn$ exhibits a series of magnetic transitions accompanying with large magnetic anisotropy and electric transport properties. Compared with the *ab*-plane easy axis in $Mn_3Sn$, it switches to the *c*-axis in $Mn_{2.42}Fe_{0.58}Sn$, producing a sufficiently large uniaxial anisotropy. At 2 K, a giant out-of-plane coercivity ($H_c$) up to 3 T was observed, which originates from the large uniaxial magnetocrystalline anisotropy. The modified Sucksmith-Thompson method was used to determine the values of the second-order and the fourth-order magnetocrystalline anisotropy constants $K_1$ and $K_2$, resulting in values of $6.0 \times 10^4$ J/m$^3$ and $4.1 \times 10^5$ J/m$^3$ at 2 K, respectively. Even though the Curie temperature ($T_C$) of 200 K for $Mn_{2.42}Fe_{0.58}Sn$ is not high enough for direct application, our research presents a valuable case study of a typical uniaxial anisotropy material.

Keywords: non-collinear antiferromagnet; coercivity; perpendicular magnetic anisotropy

# 1. Introduction

Magnetic materials possessing high $H_c$ and perpendicular magnetic anisotropy (PMA) demonstrate great potential for a range of applications such as ultrahigh-density perpendicular magnetic recording media, high-performance permanent magnets, and spintronic devices [1–6]. Particularly, large PMA is advantageous for achieving low energy consumption and high thermal stability for high-density information storage [7–9]. On the other hand, rare earth permanent magnets are currently the highest-performing permanent magnet materials as known. However, the reserves of rare earth elements are extremely limited, and the costs of extraction and processing are high. Therefore, there is an urgent need to find high-performance permanent magnets without rare earth elements to replace the widely used rare earth magnets.

[*] Corresponding author, Email: guixincao@shu.edu.cn

In recent years, $Mn_3Sn$, with hexagonal structure (Fig. 1(a)) and a Néel temperature of 420 K, has been widely studied for its excellent properties, such as the large anomalous Hall effect (AHE) [10], topological Hall effect (THE) [11], anomalous Nernst effect [12], and exchange bias effect [13]. The Mn atoms form a 2D Kagome lattice on the *ab* plane with Sn atoms sitting at the hexagon centers (Fig. 1(b)), and two neighboring Kagome $Mn_3Sn$ layers are stacked along the *c* axis (Fig. 1(a)) [14]. $Mn_3Sn$ is a room temperature non-collinear antiferromagnet, while $Fe_3Sn$, a compound with the same crystalline structure as that of $Mn_3Sn$, is a ferromagnetic metal with small coercivity and an easy axis of magnetization in the *ab* plane that can be shifted to the *c*-axis with doping [15,16]. It was reported that substituted Fe atoms at Mn sites of $Mn_3Sn$ disrupt potentially the symmetry of the Kagome lattice [17], consequently affecting the nearest-neighbor Heisenberg exchange, and anisotropic energy et al. Therefore, substituting Fe atoms at Mn sites of $Mn_3Sn$ to enhance the magnetic properties of $Mn_3Sn$ is greatly expected as Fe substitution introduces ferromagnetism to the system.

In this letter, we investigated the magnetic phase transitions of $Mn_{2.42}Fe_{0.58}Sn$ single crystals and observed a giant out-of-plane $H_c$ up to 3 T at 2 K. We find that doping Fe into $Mn_3Sn$ causes the easy magnetization axis of $Mn_3Sn$ to switch from the *ab*-plane to the *c*-axis, producing a sufficiently large uniaxial anisotropy in $Mn_{2.42}Fe_{0.58}Sn$. We use the modified Sucksmith-Thompson method which takes into account magnetization anisotropy and high-field susceptibilities to obtain magnetocrystalline anisotropy constants $K_1$ and $K_2$, resulting in values of $6.0 \times 10^4$ J/m$^3$ and $4.1 \times 10^5$ J/m$^3$ at 2 K, respectively. Both $H_c$ and $K_2$ increase with decreasing temperature ($T$), while $K_1$ is non-monotonically independent of $T$. Our result presents a typical uniaxial anisotropy system with potential to obtain the large magnetocrystalline anisotropy through Fe-doping.

## 2. Experimental details

Mn$_{2.42}$Fe$_{0.58}$Sn single crystals were synthesized by the Sn-flux method with a molar ratio of Mn: Fe :Sn =7:1.5:3 [18,19]. The obtained single crystals were shiny and hexagonal-rod-shape with typical dimensions of 2×1× 1mm³. Elemental compositions of the crystals were calculated to be Mn$_{2.42}$Fe$_{0.58}$Sn by using an energy-dispersive x-ray spectroscopy (EDS, HGST FlexSEM-1000) technique. The crystal structural characterization was performed by x-ray diffraction (XRD) using a Bruker D2 x-ray instrument with Cu Kα radiation (λ = 1.541 Å). Magnetization measurements were performed in a SQUID magnetometer (MPMS, Quantum Design). The *x*, *y*, and *z* axes that we define correspond to the [2$\bar{1}\bar{1}$0], [01$\bar{1}$0], and [0001] directions, respectively, as shown in the inset of Fig.2(d). The longitudinal resistivity was measured by a standard four-probe method using a physical property measurement system (PPMS-14, Quantum Design).

## 3. Results and discussion

The powder XRD pattern at room temperature of the Mn$_{2.42}$Fe$_{0.58}$Sn (Fig. 1(c)) shows that all the diffraction peaks of the sample can be indexed as the D0$_{19}$ type hexagonal structure with space group P6$_3$/mmc. The peaks of Mn$_{2.42}$Fe$_{0.58}$Sn shifts to right compared to the reflection peaks of Mn$_3$Sn due to the smaller atomic radius of the Fe substituting for Mn. Laue pattern of a polished (0 0 0 1) plane displays good symmetry and sharpness of the diffraction spots as shown Fig.1(d), which proves the good orientation and high quality of our single crystals.

Fig. 2(a) shows the temperature dependence of the magnetization $M(T)$ of Mn$_{2.42}$Fe$_{0.58}$Sn measured by applying a magnetic field of $\mu_0H$= 0.02 T perpendicular ($H\perp z$) and parallel ($H // z$) to the *z*-axis, respectively. It can be seen that both the $M(T)$ curves for $H\perp z$ and $H/\!/z$ under low field of $\mu_0H=0.02$ T displays a bifurcation at almost the same temperature at $T_p$ for ZFC and FC modes. Above $T_p$, there appear two transitions of $T_N \sim 297$ K (antiferromagnetic) and $T_C \sim 200$ K (ferromagnetic-like) for both $H\perp z$ and $H/\!/z$, respectively. However, when T < $T_C$, two additional transitions $T_p \sim 187$ K and $T_t \sim 163$ K emerge in the M(T) curve for $H/\!/z$, with their absence along $H\perp z$. The $T_t \sim 163$ K appeared only along $H/\!/z$ is regarded as the rotation of magnetic moments (spin reorientation) due to the enhancement of magnetocrystalline anisotropy with decreasing temperature [20,21].

This obviously different magnetic properties for $H\perp z$ and $H/\!/z$ indicate the large magnetic anisotropy in $Mn_{2.42}Fe_{0.58}Sn$. The large anisotropy is also reflected in the M(T) curves with applied various magnetic field as displayed in Fig.2(b) and (c), respectively. For $H\perp z$, the M(T) under 0.02 T shows typical ferromagnetic (FC) character with a clear FC transition at $T_C$ and complete reproduces for both ZFC and FC under 1T field. In comparison, the applied 1 T field for the M(T) curves along $H/\!/z$ causes a clear shift of $T_p$ to low temperature (low-T) from ~187 K of $\mu_0H=0.02$ T to ~50 K of $\mu_0H=1$ T, with a antiferromagnetic (AFM) character. The ZFC and FC M(T) curves were completely reproduced by applied 5 T field with a FM character as displayed in Fig.2(b). This phenomenon implies that the low-T phase of $Mn_{2.42}Fe_{0.58}Sn$ is not a completely single magnetic phase in nature [22], it can be concluded that the ground state of the sample is the coexistence of FM and AFM phases, consistent with previous reports [20,23].

Consistent with the strong magnetic anisotropy in the $Mn_{2.42}Fe_{0.58}Sn$, the large anisotropy is also reflected in the electrical transports as shown in Fig.2(d), which depict zero-field in-plane ($\rho_{xx}$) and out-of-plane resistivity ($\rho_{zz}$) between 2 and 300 K, respectively. It can be seen that $\rho_{xx}$ increases with decreasing temperature, displaying a typical semiconducting behavior. However, $\rho_{zz}$ exhibits a standard insulator-metal transition at ~200 K, corresponding to the FM transition $T_C$ in the M(T) curves. Upon further cooling from 200 K, there emerges a second transition with a typical upturn at around 40 K. These differences between $\rho_{xx}$ and $\rho_{zz}$ in $Mn_{2.42}Fe_{0.58}Sn$ are apparently different from the metallic characteristics and isotropic electrical resistivity of $Mn_3Sn$ [24]. This further displays the strong electrical anisotropy due to the Fe doping.

To further understand the large anisotropy in both the magnetic and transport properties, we measured the magnetization isotherms $M(H)$ at various temperatures for both $H\perp z$ (Fig. 3(a)) and $H//z$ (Fig. 3(b)). Interestingly, corresponding to the magnetic transitions $T_N$ and $T_C$ etc. observed in the $M(T)$ curves, the $M(H)$ curves also display varying features with temperatures. When the temperature is higher than $T_N$, i.e., 300 K, a linear behavior of $M$ with $H$ for both $H\perp z$ and $H//z$ was observed (see Fig. 3(a), 3(b)), indicating the paramagnetic phase. As the temperature is lower than $T_N$ but higher than 200 K, i.e., 250 K, a significant hysteresis was observed for $H\perp z$ as shown in inset of Fig. 3(a), suggesting the weak ferromagnetic property. The observation of weak ferromagnetic is arising from the uncompensated moment in triangular antiferromagnetic structure, consistent with that in $Mn_{2.5}Fe_{0.5}Sn$ [25]. Below $T_C$ = 200 K, the hysteresis loops exhibit typical ferromagnetic characteristics, suggesting that there is a magnetic transition from the antiferromagnetic to ferromagnetic state. In particular, an obvious hysteresis has been found in the magnetization curve at 2 K. The coercivity increase from 0 T at 200 K to 3 T at 2 K, indicating the strong magnetocrystalline anisotropy at low temperatures as shown in Fig. 3(b), consistent with the analysis of $M(T)$ curves above. Meantime, it is noteworthy that the $M(H)$ curves below 50 K appears the "collapse" phenomenon at the small field region, which is related to the weak exchange-coupling between two magnetic phases [26,27]. This evidence favors the coexistence of FM and AFM at low temperatures, in accord with the analysis in $M(T)$ curves above. In addition, the large $H_c$ and a non-saturation magnetization until 7 T at 2 K also confirm the coexistence of FM and AFM phases. To make a clear comparison, $M(H)$ curves of $Mn_{2.42}Fe_{0.58}Sn$ for both 300 K and 2 K under $H\perp z$ and $H//z$ are displayed in Figs. 3(c) and 3(d), respectively. It can be seen that the easy axis of $Mn_{2.42}Fe_{0.58}Sn$ shifts from in-plane at 300 K to out-of-plane at 2 K, along the crystallographic $c$-axis, which facilitates the usage in perpendicular magnetic recording applications. At 2 K, the out-of-plane $H_c$ is up to 3 T, but the in-plane $M(H)$ curve exhibits almost anhysteretic loop, zero remnant magnetization and high saturation field exceeding 7 T, which further realizes a large perpendicular magnetic anisotropy.

To further investigate the large out-of-plane $H_c$ and perpendicular magnetic anisotropy, we plotted $H_c$ as a function of temperature for $H \mathbin{/\mkern-6mu/} z$ orientation in Fig. 4(a). It can be seen that $H_c$ increases with decreasing temperature, and the maximum value reaches 3 T at 2 K. Considering the origin of the $H_c$, we know that the formation of hysteresis during the demagnetization process is related to the irreversibility of $M$ caused by stress fluctuations, impurities, and general magnetic anisotropy present in ferromagnetic materials [28]. The demagnetization process is caused by both the displacement of magnetic domain walls and domain wall rotation [29]. The larger hysteresis means that the domain wall displacement and/or rotation are prevented during the demagnetization process. To make this clear, we have fitted the $H_c(T)$ according to both weak domain wall pinning model $H_c \propto T$ and strong domain wall pinning mechanism $H_c^{1/2} \propto T^{2/3}$, respectively [30]. We found that our results don't agree with both models, this indicates that basically our large $H_c$ is not through the domain wall displacement. Whereas, the $H_c(T)$ curve agree well with $\log_{10} H_c \propto T$ as shown in the inset of Fig.4(a). This display that the giant $H_c$ observed in our $Mn_{2.42}Fe_{0.58}Sn$ single crystals is mainly related to the large magnetocrystalline anisotropy as it shown in $Mn_2LiReO_6$ compound [31]. In this case, the demagnetization process in our compound should be mainly achieved through the rotation of the magnetization vector. Therefore, we can conclude that the magnetic crystalline anisotropy is the main cause of the large coercivity in our compound.

Further, we estimate the magnetocrystalline anisotropy constants by using the modified Sucksmith-Thompson (ST) method. The anisotropy energy for hexagonal crystals is represented by the phenomenological expression:

$$E_a = K_1 Sin^2\theta + K_2 Sin^4\theta + K_3 Sin^6\theta \ldots (1)$$

where $K_1$ and $K_2$ are magnetocrystalline anisotropy constants and $\theta$ is an angle between the magnetization $M$ and $c$ axis. The values of $K_1$ and $K_2$ generally can be evaluated by applying the Sucksmith-Thompson (ST) method [32]. However, our magnetization measurements revealed different magnetization values above the anisotropy field as shown in Fig.3(d). This phenomenon is called magnetization anisotropy and is determined by the parameter $p = \frac{M_{S-EA} - M_{S-HA}}{M_{EA}}$, where $M_{S-EA}$ and $M_{S-EH}$ are values of the saturation magnetization for field applied along the easy and hard magnetization directions, respectively [33]. $p$ affects values of magnetocrystalline anisotropy constants calculated using the ST method. Therefore, we use the modified ST method [34,35] which takes into account magnetization anisotropy and high-field susceptibilities to obtain magnetocrystalline anisotropy constants $K_1$ and $K_2$. According to this method, the dependence of $h(1 - 4\mu p^*)$ on $\mu$ should be plotted instead of the classical one $\mu_0 H_i / M (M^2)$, where $h = \mu_0 H_i (M_{S-EA} + \chi_{EA} \mu_0 H_i)^2 / 2 M_{S-EA}, \mu = [M/(M_{S-EA} + \chi_{EA} \mu_0 H_i)]^2$ and $p^* = [p M_{S-EA} - (\chi_{HA} - \chi_{EA}) \mu_0 H_i] / [M_{S-EA} + \chi_{EA} \mu_0 H_i]$. Where $u_0 H_i$ is the internal magnetic field defined as $\mu_0 H_i = \mu_0 H - 4\pi N M_v$. Here, $M_v$ is the magnetization per unit volume and the multiplier $4\pi$ is necessary for cgs units. Since the geometric shape of the measured samples is processed as a rectangle, the demagnetization factors are estimated to be $N_x = 0.85 (H \perp z)$ and $N_z = 0.8 (H // z)$. The case of easy axis magnetic anisotropy satisfies the equation:

$$K_1 + 2K_2 \mu = h(1 - 4\mu p^*) \qquad (2)$$

The $M(H)$ curve measured along the hard axis demonstrates nonlinear behavior in Fig.3(a), which indicates the significant role of the anisotropy constant $K_2$ in the magnetization processes [36]. We obtained the saturation magnetization, $M_{S-EA}$, $M_{S-HA}$, for these two axes by extrapolating the magnetization from high fields to zero field using a linear function. The slope of this line corresponds to the high-field susceptibility, $\chi_{EA}$, $\chi_{HA}$. Field dependence of magnetization after high field susceptibility subtraction correction for $Mn_{2.42}Fe_{0.58}Sn$ single crystal at 150 K is presented in the inset of Fig. 4(b). The anisotropy of magnetization was clearly observed as $M_{S-EA} = 412$ emu/cm$^3$ and $M_{S-HA} = 291$ emu/cm$^3$ at 150 K. Using magnetization curves measured between 2 K and 150 K and taking into account the effect of the demagnetizing field, we obtained temperature dependence of the saturation magnetization $M_s$ (Fig. 4(b)), high-field susceptibility $\chi$ (Fig. 4(c)) along EA and HA for $Mn_{2.42}Fe_{0.58}Sn$, respectively. The solid lines in Fig. 4(b) is a fit of experimental $M_s(T)$ data for both $H//Z$ and $H\perp z$ using the expression [37] $M_s(T) = M_s(0)[1 - s\left(\frac{T}{T_C}\right)^{\frac{3}{2}} - (1-s)\left(\frac{T}{T_C}\right)^{\frac{5}{2}}]^{1/3}$. Here, $M_s(0)$ is a saturation magnetization at 0 K, $T_C$ is a Curie temperature and $s$ is a shape parameter, which relates to spin-wave stiffness. According to the fitting, $T_C$ of the $Mn_{2.42}Fe_{0.58}Sn$ single crystal equals to 200 K independently of crystallographic direction, which is consistent with that from $M(T)$ data. The other parameters are listed in Table 1.

The high-field susceptibility in both direction increases with increasing temperature and the ratio $\chi_{HA} > \chi_{EA}$ is typical in Fig. 4(c), which is associated with thermal fluctuations [34,35]. Using $M_s(T)$ values, we obtained the $p$ versus $T$ as displayed in the inset of Fig. 4(c). It can be seen that with increasing temperature, $p(T)$ increases monotonically. Based on above, the anisotropy constants, $K_1$ and $K_2$ for $Mn_{2.42}Fe_{0.58}Sn$ single crystal were determined by using the modified ST method. Thus, a linear approximation of $h(1-4\mu p^*)$ versus $\mu$ provides $K_1$ and $K_2$ values at 150 K in the inset of Fig. 4(d). In this way, the $K_1$ and $K_2$ at various temperatures are obtained and presented in Fig. 4(d). Within 2-150 K, both $K_1$ and $K_2$ are positive, which is consistent with the fact that $Mn_{2.42}Fe_{0.58}Sn$ has an easy magnetization axis in this temperature region. The temperature dependence of $K_1$ is non-monotonic, with a minimum at 2 K and a maximum of $0.9\times10^5$ J/m$^3$ at 100 K. It turns out that $K_2$ is an order of magnitude larger than $K_1$. $K_2$ increases monotonically with decreasing temperature, with similar trend as the temperature dependence of $H_c$. Therefore, both $H_c$ and $K_2$ reaches the maximum value at 2 K with $H_c = 3$ T and $K_2 = 4.1\times10^5$ J/m$^3$. Our results indicates that the uniaxial anisotropy of $Mn_{2.42}Fe_{0.58}Sn$ is comparable with the reported potential rare-earth free permanent magnet materials, such as MnAl ($1.67\times10^6$ J/m$^3$) [38], MnBi ($2.23\times10^6$ J/m$^3$) [39], CoPt ($1.73\times10^6$ J/m$^3$) [40] and Fe$_2$P ($2.32\times10^6$ J/m$^3$) [41].

## 4. Conclusions

In summary, we found a giant out-of-plane coercivity (3 T) in high quality $Mn_{2.42}Fe_{0.58}Sn$ single crystals with a typical easy axis along $z$ direction due to the Fe doping. This unusually large coercivity originates from the perpendicular magnetocrystalline anisotropy. The modified Sucksmith-Thompson method, which takes into account magnetization anisotropy and high-field susceptibilities, has been used to obtain the magnetocrystalline anisotropy constants $K_1$ and $K_2$. At 2 K, $K_1$ and $K_2$ result in values of $6.0 \times 10^4$ J/m$^3$ and $4.1 \times 10^5$ J/m$^3$, respectively. We found that both the values of $H_c$ and $K_2$ increase with decreasing temperature $T$, while $K_1$ is non-monotonically independent of $T$. Though the $T_C$ of 200 K for Mn$_{2.42}$Fe$_{0.58}$Sn is not high enough for direct application, our study provides a model to obtain the perpendicular magnetocrystalline anisotropy through Fe-doping in the Mn$_3$Sn compound.

**ACKNOWLEDGMENT**

This work was supported by Key Research Project of Zhejiang Lab (No. 2021PE0AC02), Department of Science and Technology of Zhejiang Province (2023C01182) and Shanghai Engineering Research Center for Integrated Circuits and Advanced Display Materials. A portion of this work was performed on the Steady High Magnetic Field Facilities, High Magnetic Field Laboratory, CAS.

**Figure Captions:**

**FIG. 1.** (a) Crystal structure of Mn$_3$Sn. (b) Triangular spin structure in the Kagome layer (*ab* plane). (c) Powder XRD data taken from the crushed single crystals of Mn$_{2.42}$Fe$_{0.58}$Sn. Inset shows photographic image of as-grown Mn$_{2.42}$Fe$_{0.58}$Sn single crystal. (d) Laue pattern of a polished surface of Mn$_{2.42}$Fe$_{0.58}$Sn single crystal showing the (0 0 0 1) plane.

**FIG. 2.** (a) $M(T)$ curves under the field of 200 Oe applied both in parallel and perpendicular to the *z*-axis for field-cooled (FC) and Zero-field-cooled (ZFC) modes, respectively. (b) and (c) show $M(T)$ curves measured in the FC and ZFC modes under various field in parallel and perpendicular to the *z*-axis, respectively. The data in ($\mu_0 H$ = 0.02 T) is shifted along the y axis for better clarity. (d) $\rho_{xx}$(T) (left) *and* $\rho_{zz}$(T) (right) curves for Mn$_{2.42}$Fe$_{0.58}$Sn.

**FIG. 3.** Magnetization isotherms, $M(H)$ of Mn$_{2.42}$Fe$_{0.58}$Sn for $H \perp z$ (a) and $H // z$ (b) at

various temperatures; Inset shows enlarged curves of central region of the hysteresis loops for 250 K. (c) and (d) show $M(H)$ isotherms of $Mn_{2.42}Fe_{0.58}Sn$ for $H \perp z$ and $H /\!/ z$ at 300 K and 2 K, respectively.

**FIG. 4.** (a) Coercivity as a function of temperature for $H /\!/ z$. Inset shows the fitted logarithmic dependence of $H_c$ with temperature. (b)Temperature dependences of $M_s$; Inset shows the field dependence of magnetization after high field susceptibility subtraction correction for $Mn_{2.42}Fe_{0.58}Sn$ single crystal at 150 K. (c)Temperature dependences of the high-field susceptibility, $\chi$; Inset shows the temperature dependence of magnetization anisotropy $p$. (d) Temperature dependencies of magnetocrystalline anisotropy constant $K_1$ and $K_2$ of the $Mn_{2.42}Fe_{0.58}Sn$ single crystal obtained by the modified Sucksmith-Thompson method; Inset shows the application of the method.

**Table 1**

The Curie temperature $T_C$, saturation magnetizations $M_s$ at 0 K and shape parameters $s$ along hard magnetization axis (HA) and easy magnetization axis (EA) of the $Mn_{2.42}Fe_{0.58}Sn$ single crystal.

| Direction | $T_C$ (K) | $M_s$ (0) (emu/cm$^3$) | $s$ |
|---|---|---|---|
| HA | 200 | 463 | 1.53 |
| EA | 200 | 573 | 0.78 |

Fig.1 W.-H Shen et al.

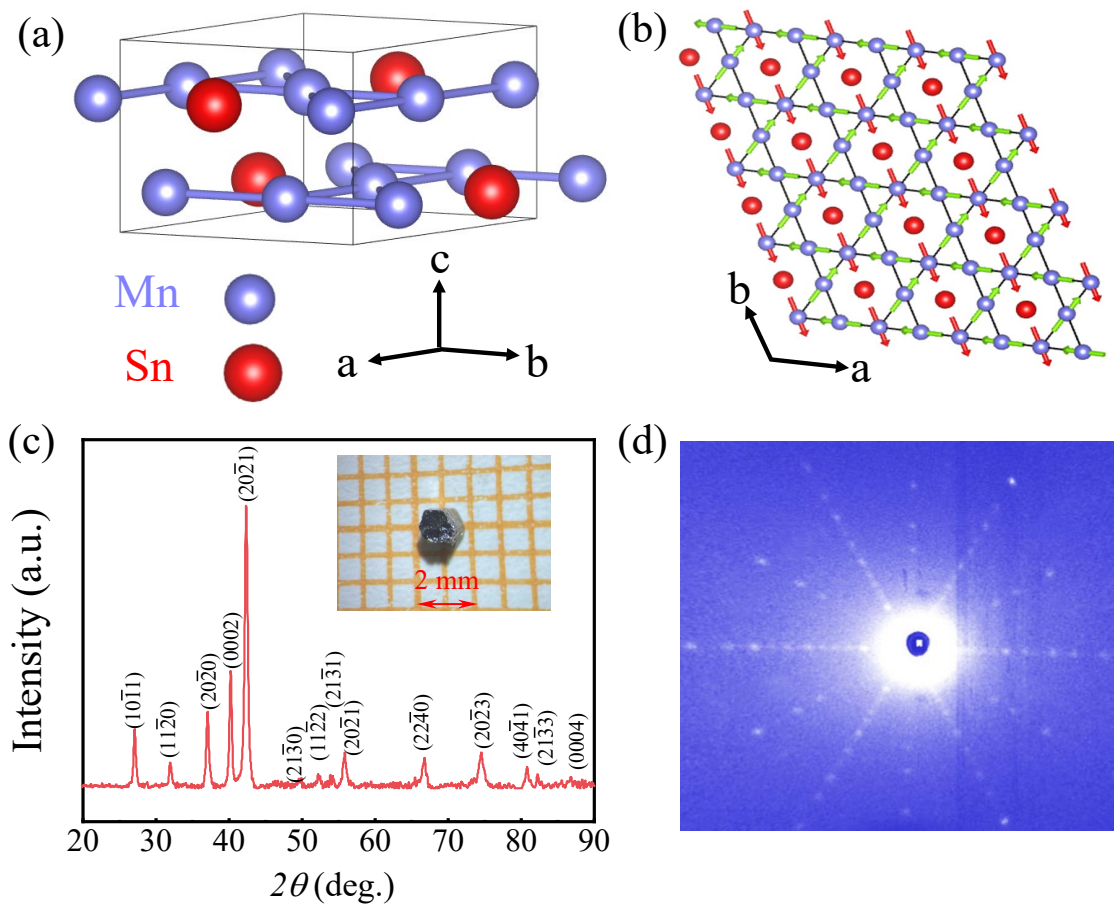

Fig.2 W.-H Shen et al.

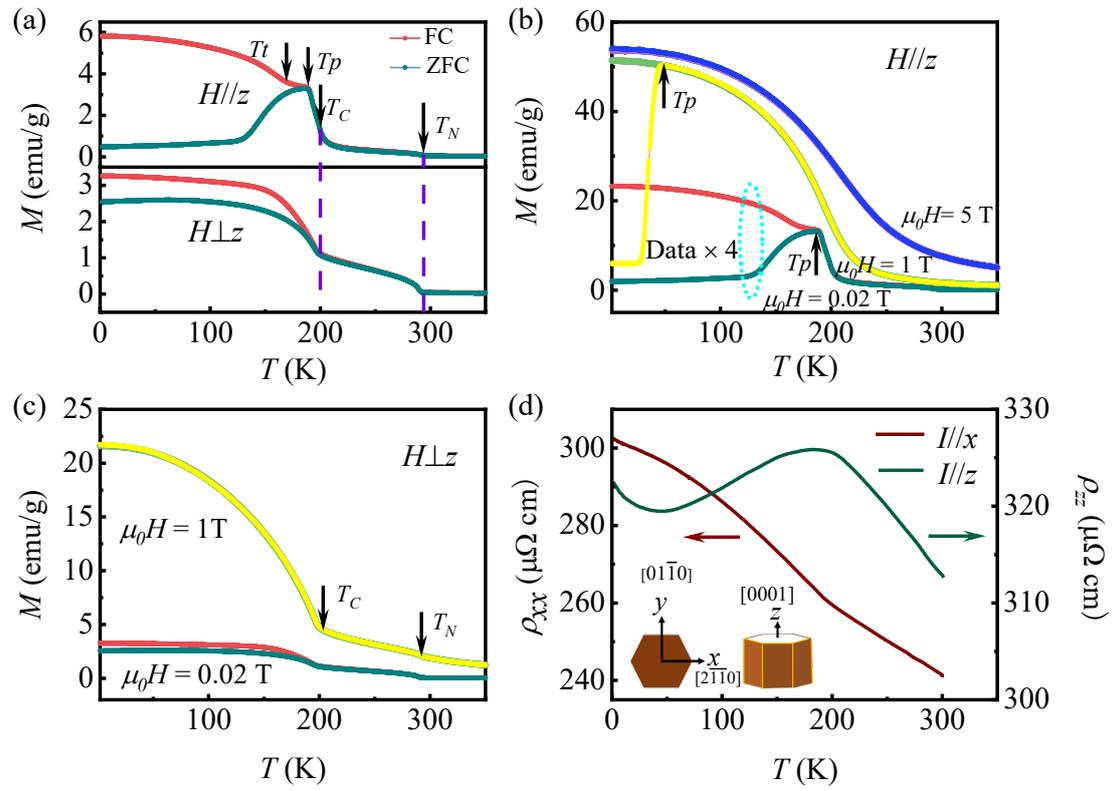

Fig.3 W.-H Shen et al.

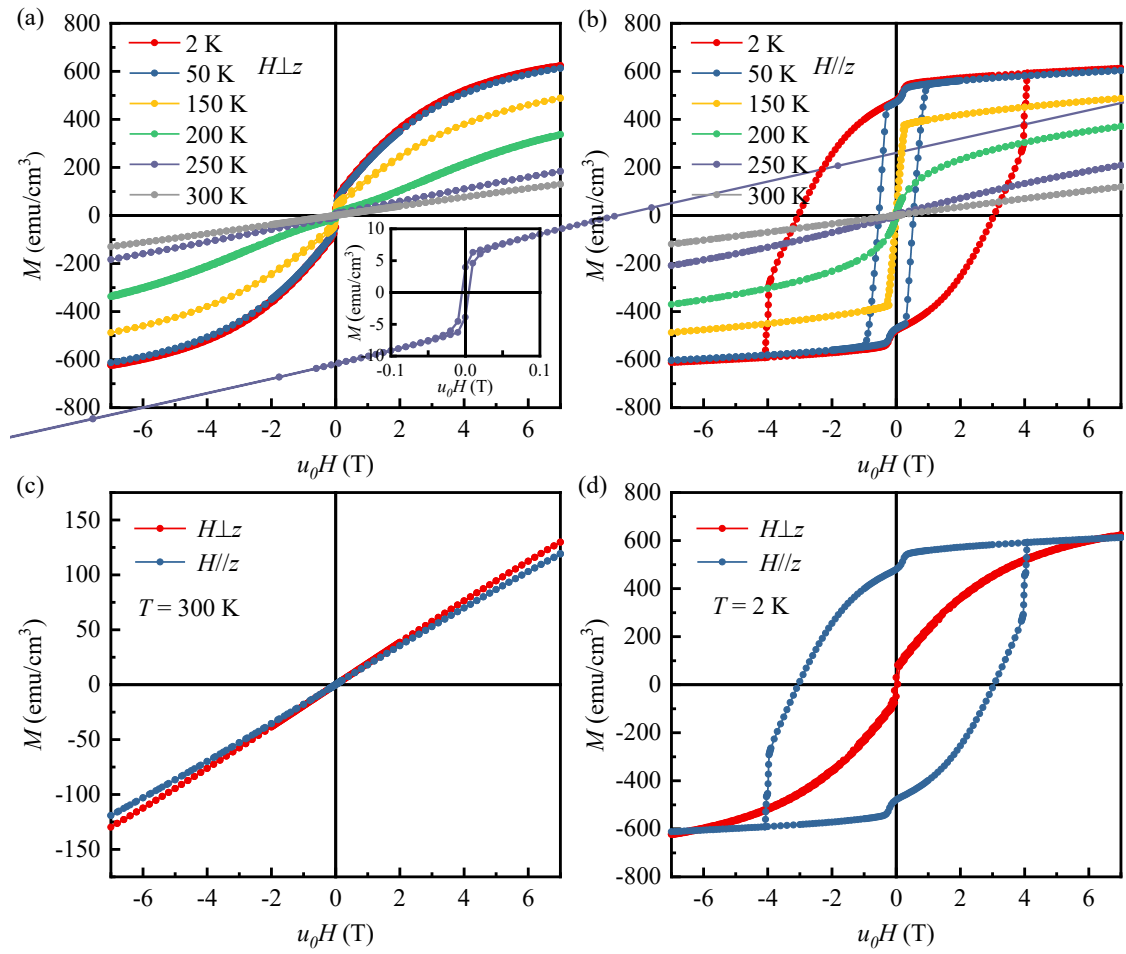

Fig.4 W.-H Shen et al.

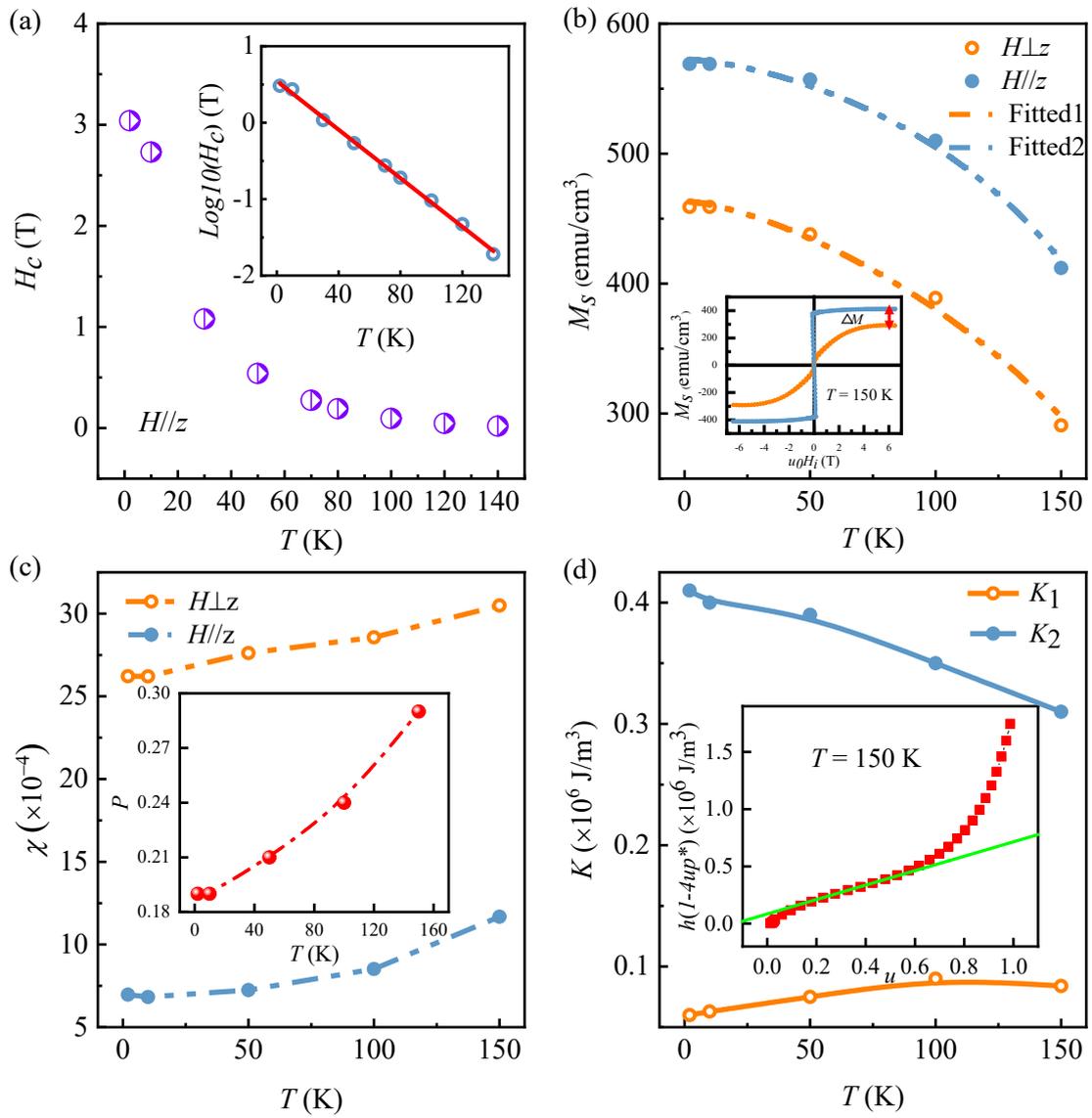